# *Measurements of spin polarization of FeB and FeCoB nanomagnets by the Anomalous Hall effect*

*Vadym Zayets*
Spintronics Research Center,
National Institute of Advanced Industrial Science and Technology (AIST), Tsukuba, Japan

Abstract:
*A new method to measure the spin polarization of a ferromagnetic nanomagnet is proposed and experimentally demonstrated. A high precision, repeatability and simplicity are the features of the proposed method. The dependency of the spin polarization on the gate voltage and the bias current were observed.*

## 1. Introduction

The spin polarization of the electron gas of conduction electrons is an important parameter, which determined the magnetic and magneto-transport properties of a metal. The spin polarization in a metal can be measured by the Andreev reflection[1,2], spin-dependent photoluminescence[3], Raman spectroscopy[3] and spin LED[4]. It can be estimated from the tunnel magneto-resistance using the Julliere formula[5]. The measurement of the spin polarization is a challenging task. The difficulties of such measurement and the discrepancies between the data from measurements by different methods are discussed in Ref.[6].

The spin polarization can be measured using the Anomalous Hall effect [7,8] (AHE). Merits of such measurement are simplicity and the ability to measure the spin polarization even in the case of a nano-sized object. In Ref.[7], the spin polarization $sp$ in a MgZnO/ZnO heterostructure was estimated from a measurement of dependence of the Hall angle $\alpha_{AHE}$ of the AHE on an external magnetic field $H_{ext}$. The MgZnO is a paramagnetic material. In the absence of an external magnetic field, both the net magnetization and the spin polarization of the MgZnO are zero. As a result, in the absence of $H_{ext}$, the AHE is not observed in the MgZnO. An external magnetic field induces the magnetization in MgZnO and creates the spin polarization. As a result, in a sufficient magnetic field the AHE is observed[7] in the MgZnO. Since both the magnetization and the spin polarization in the MgZnO increase as the magnetic field increases, the Hall angle $\alpha_{AHE}$ increases as well. It is possible to evaluate both the induced magnetization and the spin polarization from measured dependence of $\alpha_{AHE}$ on the magnetic field[7].

In the case when both the magnetization and the spin polarization depend on the external field, there is some uncertainty in the estimation of the spin polarization from $\alpha_{AHE}$. Even a small deviation of either the direction of the magnetization or its absolute value from the expected values may cause a substantial error in the evaluation of the spin polarization. A high measurement precision of the $sp$ from the AHE can be expected only in the case when neither the direction nor absolute value of the magnetization depends on an external magnetic field. In this case, the dependence of the $sp$ on $H_{ext}$ is only contribution to the measured dependence of $\alpha_{AHE}$ on $H_{ext}$ and the evaluation of $sp$ is straightforward.

A nanomagnet made of a ferromagnetic metal with the perpendicular magnetic anisotropy (PMA)[9] is well-suitable for such measurement. Its magnetization is firmly fixed perpendicularly to the film surface by the strong spin-orbit interaction. Additionally, its magnetization is easily saturated even in a weak external magnetic field[9]. In this case there is no any magnetic domain in the nanomagnet and the magnetization is directed in one direction through the whole volume of the nanomagnet. Such case is the simplest for the fitting of the measured $\alpha_{AHE}$ and the evaluation of $sp$.

In the paper, a new method to measure the spin polarization in a nanomagnet made of a ferromagnetic metal by the AHE is proposed and demonstrated. The spin polarization in FeB and FeCoB nanomagnets was measured. The dependences of the spin polarization on the gate voltage and on the polarity of the bias current were observed.

## 2. Spin polarization of electron gas in a magnetic field

All conduction electrons in a ferromagnetic metal can be divided into groups of spin-polarized and spin-unpolarized electrons. In the group of the spin-polarized electrons, the spins of all electrons are in the same direction. In the group of the spin-unpolarized electrons, the spins are distributed equally in all directions[10]. Additionally, there are some electrons, which are "spin-inactive". A pair of these electrons with opposite spins occupies one quantum state.

The occupation of quantum states by the electrons of both the spin-polarized and spin-unpolarized groups is one electron per a state. As a result, the spin of each state is 1/2 and the spin direction for each quantum state is defined. The spin direction represents the direction of the local breaking of the time-inverse symmetry for the state. When a quantum state is occupied by two conduction electrons of opposite spins, the spin of such quantum state is zero. As a result, the spin direction of this state cannot be defined and the electrons occupying this state are "spin-inactive"[10]. The electrons, which energy is substantially below the Fermi energy, mainly belong to this group. For example, nearly all of the "deep level" electrons belong to this group. In contrast, the energy of electrons of the groups of spin-polarized and spin-unpolarized electrons is distributed mainly nearly the Fermi energy. The energy distributions of electrons of each group have been calculated in Ref.[11].

The spin polarization *sp* of the electron gas is defined as a ratio of the number of spin-polarized electrons to the total number of the spin-polarized and spin-unpolarized electrons

$$sp = \frac{n_{TIA}}{n_{TIA} + n_{TIS}} \quad (1)$$

where $n_{TIA}$ and $n_{TIS}$ are the numbers of spin-polarized and spin-unpolarized electrons, respectively. It should be noted that the "spin-inactive" electrons are not included in the definition of the spin polarization.

The amount of electrons in each group is determined by a balance between the spin pumping and the spin damping. The spin pumping is the conversion of electrons from groups of spin-unpolarized electrons into the group of the spin-polarized electrons. The spin damping is the conversion in the opposite direction. There are several mechanisms of the spin pumping. The first mechanism is the exchange interaction between the localized d-electrons and the delocalized conduction electrons. The second mechanism is the scattering of a localized d-electron into the state of the conduction electrons. Both these mechanisms align the spins of the conduction electrons along the spins of the localized electrons. The third mechanism of the spin pumping is the alignment of spins of conduction electrons along a magnetic field. This mechanism can be understood as follows. There is a spin precession around direction of the magnetic field. Additionally to the precession, the electron spin is aligned along the magnetic field (Fig.1(a)). The effect is called the precession damping and it is described by the Landau–Lifshitz equation. The precession damping aligns spins of the spin-polarized electrons along the direction of the magnetic field. The rate of the spin pumping in the magnetic field was calculated in Ref. [10]. In the case when the magnetic field $H$ is applied along the magnetization, the rate of the spin pumping is calculated as[10]

$$\frac{\partial n_{TIA}}{\partial t} = \frac{n_{TIS}}{t_{H,pump}} \quad (2)$$

where $t_{H,pump}$ is the effective spin pumping time.

When the spin pumping time $t_{H,pump}$ is substantially larger than the average scattering time $t_{scat}$ of a conduction electron

$$t_{H,pump} \gg t_{scat} \quad (3),$$

the spin pumping time is reverse proportional to the $H$ and it can be calculated as[10]

$$t_{H,pump} = \frac{3}{2 \cdot g \cdot \lambda \cdot \mu_B \cdot H} \quad (4)$$

where $g$ is the g-factor of a conduction electron, $\lambda$ is the is a phenomenological damping parameter of the Landau–Lifshitz equation and $\mu_B$ is the Bohr magneton.

In case of a larger magnetic field or a larger damping parameter, the condition (1.2) is not satisfied and the dependence of $t_{pump}$ on H becomes non-linear [10]. The approximation, which is used under the condition (1.3), is the assumption of a constant angular speed of the spin alignment in time between two consequent electron scatterings. All measurements, which are described in this paper, are done in a weak magnetic field, for which the condition (1.3) is assumed to be satisfied.

As was mentioned above, additionally to the spin alignment along the magnetic field, the spins of the conduction electrons align along the magnetization due to the sp-d exchange interaction and the sp-d scatterings. Similarly, this alignment can be described as a conversion of electrons from the group of the spin-polarized electrons into the group of spin-unpolarized electrons. The conversion rate can be described as

$$\frac{\partial n_{TIA}}{\partial t} = \frac{n_{TIS}}{t_{sp,pump}} \quad (5)$$

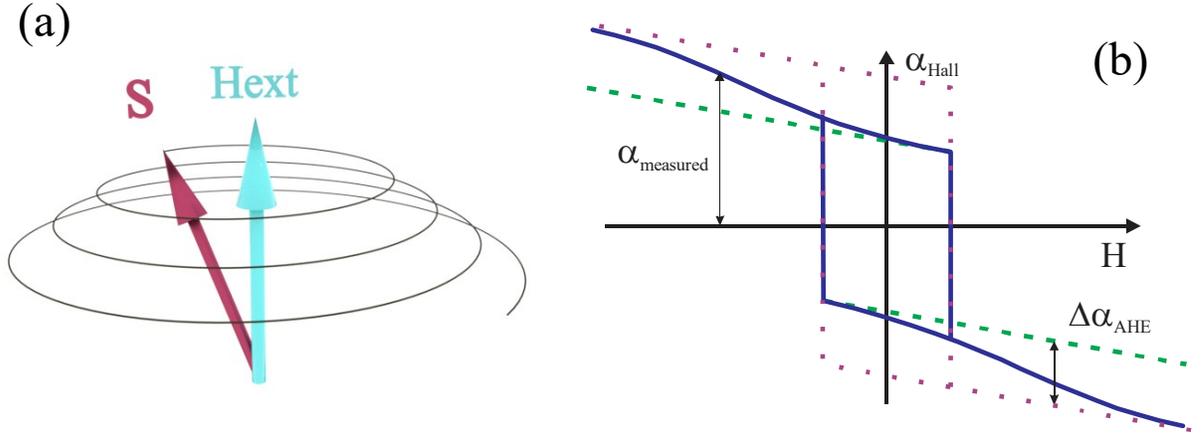

Fig. 1 . (a) Alignment of electron spin along a magnetic field due the precession damping is the origin of the spin pumping and the increase of spin polarization *sp* in the magnetic field. (b) Schematic diagram of measured hysteresis loop of Hall angle $\alpha_{Hall}$ in a ferromagnetic metal with PMA (solid blue line). The green dash line shows $\alpha_{Hall}$ in the imaginary case if *sp* would not depend on the magnetic field. The red dot line shows $\alpha_{Hall}$ in the case of a metal, in which *sp* is close to 100 %.

where $t_{sp,pump}$ is the spin pumping time due to the sp-d exchange interaction and scatterings.

The spin relaxation is a mechanism, which re-aligns the spins of spin-polarized electrons from one direction. The spin re-alignment can be described by the electron conversion from the group of the spin-polarized electron to the group of the spin-unpolarized[10–12]. The conversion rate of the spin relaxation is given as

$$\frac{\partial n_{TIS}}{\partial t} = \frac{n_{TIA}}{t_{relex}} \quad (6)$$

where $t_{sp}$ is the spin relaxation time

In equilibrium, the conversion rates between groups of spin- polarized and -unpolarized electrons should be balanced:

$$\frac{\partial n_{TIS}}{\partial t} = \frac{\partial n_{TIA}}{\partial t} \quad (7)$$

Substitution of Eqs. (1.2),(1.4) - (1.6) into Eq.(1.7) gives the ratio $n_{TIA}/n_{TIS}$. Substitution of this ratio into Eq. (1.1) gives the spin polarization of the electron gas as

$$sp = sp_0 \frac{1 + \frac{H}{H_{pump}} \frac{1-sp_0}{sp_0}}{1 + \frac{H}{H_{pump}}(1-sp_0)} \quad (8)$$

where

$$H_{pump} = \frac{3}{2 \cdot g \cdot \lambda \cdot \mu_B \cdot t_{relex}} \quad (9)$$

and $sp_0$ is the equilibrium spin polarization of the electron gas in absence of an external magnetic field and it is calculated as

$$sp_0 = \frac{t_{relex}}{t_{sp,pump} + t_{relex}} \quad (10)$$

### 3. Measurement method

The spin polarization of the electron gas was evaluated from a measurement of the dependence the $\alpha_{AHE}$ on the applied magnetic field. The proposed method uses the fact that the Hall angle $\alpha_{AHE}$ of the AHE is linearly proportional to the out-of-plane component of magnetization $M_\perp$ and the spin polarization $sp$. It can be calculated as

$$\alpha_{AHE} = \frac{\sigma_{xy}}{\sigma_{xx}} = a \cdot M_\perp \cdot sp \quad (11)$$

where $\sigma_{xx}$ and $\sigma_{xy}$ are diagonal and off-diagonal components of the conductivity tensor and *a* the proportionality constant.

The linear dependence of $\alpha_{AHE}$ on the $M_\perp$ is well verified experimentally. For example, the PMA strength of a ferromagnetic metal is measured from the dependence of the in-plane magnetization component on an applied in-plane magnetic field. The magnetization measurement by AHE and the direct magnetization measurements by a magnetometer give the exact same results. From both measurements the dependence is linear and the evaluated anisotropy field is the same for both methods[9,13,14].

The linear dependence of $\alpha_{AHE}$ on the spin polarization $sp$ can be explained as follows. The AHE is originated from the spin-dependent scatterings of the conduction electrons[13,15–21]. In a metal with a non-zero perpendicular component of magnetization the probabilities of a scattering of a conduction electron to the left and to the right are different with respect to the current direction[13,15]. Due to this difference, there are more electrons accumulated at one side of a metallic wire and the charge accumulation creates the Hall voltage. The spin-unpolarized and spin-unpolarized electrons contribute differently to the AHE[22]. For example, let us assume that there are more spin-up electrons scattered to the left and more spin-down scattered to the right. As a result, the spin-up and spin-down electrons are accumulated at left and right sides of the metallic wire, respectively. This effect of the creation of spin accumulation is called the Spin Hall effect[23,24]. In the group of spin-unpolarized electrons, the electron spins are distributed equally in all directions. Even though the scattering of the spin-up and spin-down electrons depends on the scattering direction, the number of spin-up scattered to the left equals to the number of spin-down electrons scattered to the right and vice versa. As a result, the scattering probability of all spin-unpolarized electrons is direction- independent. The scattering of spin-unpolarized electrons produces a spin accumulation, but no charge accumulation and consequently no Hall voltage. The case is different for the spin-polarized electrons. The spin direction is the same for all spin-polarized electrons and the directional difference of the scattering probability is the same for all electrons of this group. As a result, the amounts of spin-polarized electrons, which are scattered to the left and to the right, are different and a charge is accumulated at the sides of the wire. The Hall voltage and consequently the Hall angle $\alpha_{AHE}$ is linearly proportional to the number of the scattered spin-polarized electrons and therefore $\alpha_{AHE}$ is proportional to the spin polarization as it is described by Eq. (11).

The dependence of $\alpha_{AHE}$ on $sp$ is a well-verified experimental fact. For example, Wonderlich et al.[8] have studied the AHE in an AlGaAs/GaAs heterojunction. Since the AlGaAs/GaAs is a paramagnetic material, the AHE was observed in it only when the spin polarization in this material was created by circularly-polarized light. The AHE was measured by a pair of nano-sized Hall probes and the $sp$ was created by a focused laser beam[8]. The AHE was observed even when the laser beam was focused slightly away and the spin polarization was created at a distance from the Hall probe. In this case, the spin polarized electrons diffused towards the Hall probes and the electron gas still remains spin-polarized in the vicinity of the Hall probe. When the distance becomes longer, the less spin-polarized electrons reach the Hall probe and the spin polarization at the Hall probe becomes smaller. Experimentally, the reduction of $\alpha_{AHE}$ has been observed when the laser beam was moved away from the Hall probe[8]. The reduction of $\alpha_{AHE}$ in this experiment can only be explained as being caused by the reduction of $sp$. It proves the dependence of $\alpha_{AHE}$ on $sp$.

Additionally to the AHE, a conduction electron in a magnetic field experiences the ordinary Hall effect (OHE), which is induced by the Lorentz force and is linearly proportional to the perpendicular component of the magnetic field. The total measured Hall angle $\alpha_{Hall}$ is the sum of the Hall angle of the AHE and OHE. From Eqs. (10),(11) it can be calculated as

$$\alpha_{Hall} = \alpha_{AHE} + \alpha_{OHE} = \alpha_{AHE,0} \frac{1 + \frac{H_\perp}{H_{pump}} \frac{1 - sp_0}{sp_0}}{1 + \frac{H_\perp}{H_{pump}}(1 - sp_0)} + \beta_{OHE} \cdot H_\perp \quad (12)$$

where $\alpha_{AHE,0}$ is the Hall angle without an external magnetic field, $H_\perp$ is the magnetic field applied perpendicularly to the film and $\beta_{OHE}$ is the Hall coefficient. It should be noted that $\alpha_{AHE,0}$ also includes an OHE contribution induced by the demagnetization field.

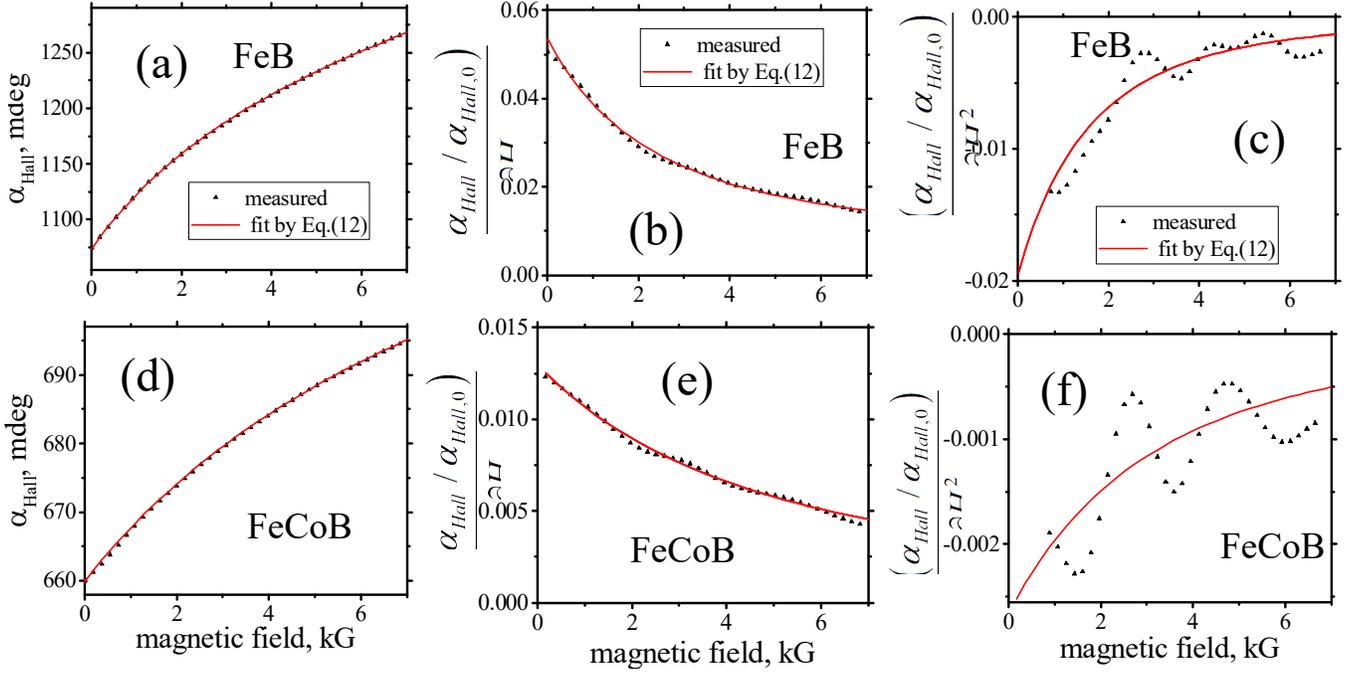

Fig. 2 . (a),(d) Hall rotation angle $\alpha_{Hall}$ as function of applied magnetic field $H$ ; (b) ,(e) 1st derivative (c),(f) 2nd derivative of $\alpha_{Hall}$ normalized to its value at H=0.   (a),(b),(c)  FeB nanowire (sp=81.7%)  (d),(e),(f) FeCoB (sp=89.4 %). Black triangles show the measured data. The red line shows the fitting by Eq.(1.12)

Figure 1(b) shows the measured dependence of the $\alpha_{Hall}$ on magnetic field (solid blue line) in a ferromagnetic metal with PMA. The green dash line shows $\alpha_{Hall}$ in the imaginary case if the spin polarization would not depend on magnetic field. In this case, $\alpha_{AHE}$ is a constant and does not depend on the magnetic field. It only reverses its sign when the magnetization direction is reversed. The $\alpha_{OHE}$ is linearly proportional to the magnetic field.  The red dot line shows $\alpha_{Hall}$ in the case of metal in which *sp* is close to 100 %, but other properties are the same. According to Eq.(11), in this case $\alpha_{AHE}$ is larger. Similarly it does not depend on the magnetic field, because the spin polarization cannot be larger than 100% and therefore it is saturated. In the realistic case (the solid line), the $\alpha_{Hall}$ increases from the dot line to the dash line as the spin polarization in the metal increases. In this paper the spin polarization is evaluated from fitting the measured dependence of Fig.1(b) by Eq.(12).

## 4. Experiment

The FeB and FeCoB samples were fabricated on a Si/SiO2 substrate by sputtering. The layer stack of the FeB sample is $SiO_2$:Ta (2.5):FeB(1):MgO(8): $SiO_2$ and  of the FeCoB sample is $SiO_2$:Ta (5):$Fe_{0.4}Co_{0.4}B_{0.2}$ (1):MgO(8): $SiO_2$, where thickness is in nanometers.   A nanowire of different width between 100 and 1000 nm with a Hall probe was fabricated by the argon milling. The width of the Hall probe is 50 nm. The FeB and FeCoB layers were etched out from top of the nanowire except a small region of the nanomagnet, which was aligned to the Hall probe. The nanomagnets of different lengths between 100 nm to 1000 nm were fabricated.

When it is not mentioned, the Hall angle is measured at current density of 5 mA/µm$^2$. The $\alpha_{AHE}$ in the ferromagnetic metal was evaluated as

$$\alpha_{AHE} = \left(1 + \frac{\sigma_{nonMag} \cdot t_{nonMag}}{\sigma_{ferro} \cdot t_{ferro}}\right) \frac{V_{Hall}}{I \cdot R} \frac{L}{w} \quad (13)$$

where  $\sigma_{ferro}$ , $\sigma_{nonMag}$ are conductivities of ferromagnetic and non-magnetic metals; $t_{ferro}$ , $t_{nonMag}$ are their thicknesses, $V_{Hall}$ is the measured Hall voltage, $I$ is the bias current and $R,L,w$ are the resistance, length and width of the nanowire, correspondently. The merit of usage of $\alpha_{Hall}$ instead of the Hall resistivity $R_{Hall}$ is that $\alpha_{Hall}$ does not depend on the device geometry. It depend only on material parameters (Eq.(11)).

Both the $\alpha_{AHE}$ and $\alpha_{OHE}$ reverse their sign, when M and H are reversed. In order to avoid a systematic error due to a possible misalignment of the Hall probe, the Hall angle was measured as

$$\alpha_{Hall} = \frac{\alpha_{Hall}(H,M) - \alpha_{Hall}(-H,-M)}{2} \quad (14)$$

Figure 2 shows the measured $\alpha_{Hall}$ and its first and second derivatives as function of applied magnetic field **H** and their simultaneous fitting by Eq.(12). For both samples, the fitting of 0, 1st and 2nd derivatives are nearly excellent. It implies that Eq.(12) well describes the magnetic-filed dependence of $\alpha_{Hall}$. The best fitting of Fig.2 was obtained for values $sp_0$=81.7% and $H_{pump}$=0.834 kG in the case of the FeB sample and $sp_0$=89.4 % and $H_{pump}$=1.098 kG in the case of the FeCoB sample. The difference of $sp_0$ is only 8% between the FeB and FeCoB samples. However, the differences of 1st and 2nd derivatives are substantial (See Fig.2). Such a high sensitivity of 1st and 2nd derivatives to a small change of $sp_0$ is the reason of a relatively high measurement precision of this method. For the measurements of Fig.2 it was estimated to be at least 0.1%.

The measured $sp_0$ was slightly different for nanomagnets of different sizes fabricated on the same wafer. On average, the $sp_0$ of the smallest samples was a few percent smaller than that of the largest samples. For devices of smaller $sp_0$, the $H_{pump}$ was always larger. It implies that the decrease of $sp_0$ in this case is due to the decrease of the spin relaxation time $t_{relax}$ (Eq.(6)), which is most likely due to the increase of the number of nanofabrication defects. The case is different for nanomagnets made of a different material. The FeCoB nanomagnets have both the larger $sp_0$ and the larger $H_{pump}$ in comparison with the FeB nanomagnets. It means that the higher $sp_0$ in FeCoB is not due to decrease of $t_{relax}$, but due to a stronger sp-d spin pumping and a larger $t_{sp,pump}$ (Eq.(5)).

It should be noted that Eq.(12) describes the bulk-type of AHE. Additionally, there is a smaller interface contribution to AHE, which is not included in Eq.(12). From our study, in case of a ferromagnetic metal thinner than ~0.2 nm the interface contribution to AHE becomes comparable with the bulk contribution and the measured $\alpha_{Hall}$ cannot be fitted by Eq.(12).

## 4. Voltage-induced magnetic anisotropy (VCMA effect)

The VCMA effect describes the fact that in a capacitor, in which one of the electrodes is made of a thin ferromagnetic metal, the magnetic properties of the ferromagnetic metal are changed, when a voltage is applied to the capacitor. For example, under an applied voltage the magnetization direction of the ferromagnetic metal may be changed [25–27] or even reversed [28,29]. This magnetization-switching mechanism can be used as a data recording method for low-power magnetic random access memory [30–32] and all metal transistor [33]. Until now the physical origin of the VCMA effect has not been clarified. However, several possible physical mechanisms have been discussed [14,34–36].

Several magnetic properties of a nanomagnet may be influenced by a gate voltage. It was observed experimentally that the anisotropic field $H_{anis}$ [37–39], coercive field $H_c$ [40–42], Hall angle $\alpha_{Hall}$ [14,37] logarithm of magnetization switching time $t_{switch}$ [14] and logarithm of the retention time $t_{reten}$ [14] depend linearly on the gate voltage. The observed polarity of the voltage-dependence is the same in all cases. The $H_{anis}$, $H_c$, $\alpha_{Hall}$, $t_{switch}$, and $t_{reten}$ become larger at a negative gate voltage and smaller at a positive gate voltage [14].

Here we report measurements of the modulation of the spin polarization of a nanomagnet by a gate voltage. Figure 3(a) shows the structure of the studied device. The device design is the same as in Ref.[14]. The Ta(1):Ru(5) gate electrode was deposited on top of the MgO gate for the FeB and FeCoB sample. The gate voltage is applied between the gate electrode and one end of the nanowire.

Figure 3(b) show the gate-voltage dependence of the spin polarization in the FeB sample. The measured voltage-dependent change of the spin polarization is -0.245 %/V. The change is substantial and it can be reliably measured. For example, the observed voltage-dependence of 1st derivative of $\alpha_{Hall}$ (Fig.3(c)) is about 5 %. For the FeCoB sample the voltage-dependence of *sp* was smaller ~ -0.0068%/V. For both samples, the *sp* depends linearly on the gate voltage and the polarity of the dependence is the same as the polarity of the voltage-dependence of $H_{anis}$, $H_c$, $\alpha_{Hall}$, $t_{switch}$, and $t_{reten}$.

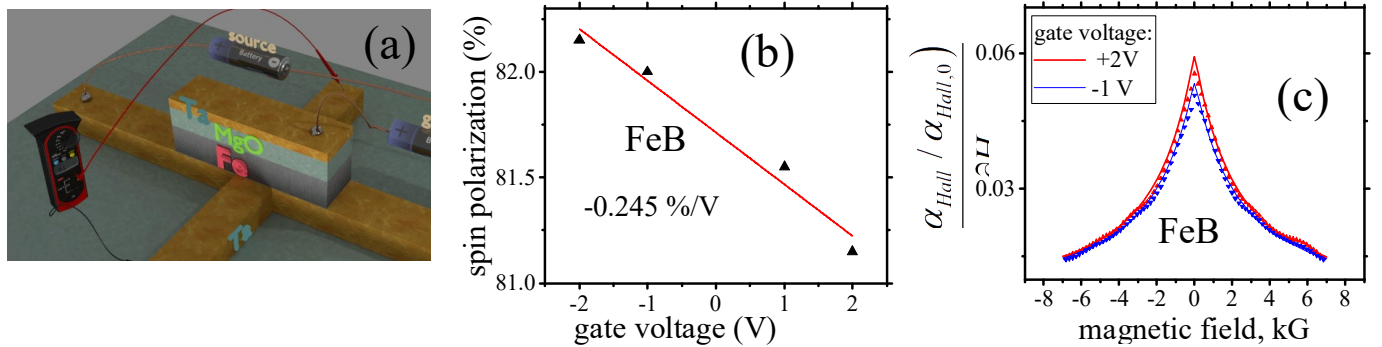

Fig. 3. VCMA effect. (a) top-view of measurement setup; (b) spin polarization of FeB nanowire as function of gate voltage; (c) 1st derivative of normalized Hall rotation angle at different gate voltage.

From Eq.(11), the value $\alpha_{Hall}/sp$ depends only on the magnetization, but not on the spin polarization. Within an experimental error, we have not detected a dependence of the value $\alpha_{Hall}/sp$ on the gate voltage. It means that the observed voltage-dependence of $\alpha_{Hall}$[14] is mainly due to the voltage-dependence of *sp*, but not magnetization. This experimental fact can be understood as follows. As was assumed in Ref. [14], a negative gate voltage enlarges the metal magnetization in a very thin region close to the gate interface. Due to the larger magnetization, the *sp-d* spin pumping (Eq.5) should be enhanced. From Eq. (10), such enhancement leads to a larger $sp_0$. The spin polarization *sp* is a feature of non-localized conduction electrons. In contrast to the localized change of the magnetization, the change of *sp* may spread deeply through the film thickness due to the diffusion of the spin-polarized electrons. In the case of a relatively thick film as was studied here, the bulk of the metal mostly contributes to the $\alpha_{Hall}$. The contribution of a thin interface region, where the magnetization is change, may be negligibly small. That is a possible reason why the gate-voltage dependence of $\alpha_{Hall}$ is mainly due to the gate-voltage dependence of *sp* and why the voltage-dependent magnetization change cannot be detected by this measurements.

The observed larger voltage-dependence of *sp* in the FeB sample comparing to the FeCoB sample, is well fitted to the explanation above. The equilibrium spin polarization $sp_0$ is larger in FeCoB sample than in FeB sample. As a result, the same change of the *sp-d* spin pumping (Eq.(5)) causes a smaller change of $sp_0$ in the FeCoB sample than in the FeB sample.

## 5. Spin-orbit torque (SOT effect)

The SOT effect describes the fact that magnetic properties of ferromagnetic nanowire may depend on the magnitude and polarity of an electrical current flowing through the nanowire. For example, under a sufficiently large current the magnetization of the nanowire may be reversed[43,44]. The direction of the magnetization reversal depends on the polarity of the current. The effect may be used as a recording mechanism for 3-terminal MRAM[45]. The origin of the SOT effect is the spin Hall effect[23,24], which describes the fact that an electrical current may create a spin accumulation.

Figure 4 (a) shows the measured spin polarization *sp* of the FeB sample as a function of the current density. The *sp* decreases for both polarities of the current. The used current density is relatively large and the decrease of *sp* is assumed to be due to the heating of the nanomagnet. The increase of the nanowire resistance confirms the increase of the nanowire temperature. In order to exclude the influence of heating, the spin polarization was measured at the same current, but for two opposite current directions. Figure 4(b) shows the change of the spin polarization as the polarity is reversed. The change of the *sp* linearly depends on the current. The measured slope is +0.00524 %/(mA/μm$^2$) for the FeB sample and -0.018 %/(mA/μm$^2$) for the FeCoB sample.

It is known that the Spin Hall effect generates spin polarization of different polarities at opposite sides of nanomagnet[23,24]. In the case of a symmetrical nanomagnet, the generated spin polarization is the same at opposite sides of the nanomagnet, the total generated spin polarization is zero and there is no SOT effect. Our studied samples are asymmetric. The ferromagnetic metal is contacting the MgO at one side and the Ta at another side. As a result, the total spin polarization generated by the Spin Hall effect is non-zero. However, in this case the contributions from each interface are nearly equal. It can explain the observed substantial change of the slope for different nanomagnets on the same wafer and the different slope polarities for the FeB and FeCoB samples. The enlargement of structure asymmetry[46,47] and optimizing interfaces may increase the current-induced change of *sp*.

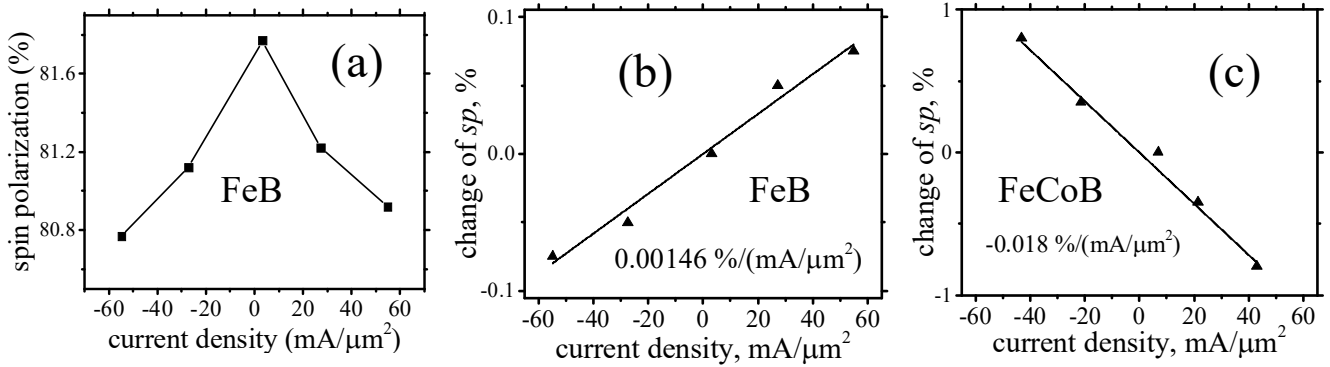

Fig. 4 . SOT effect. (a) spin polarization of FeB nanowire as function of current density. Change of the spin polarization under reversal of current direction (b) FeB (c) FeCoB

## 6. Conclusion

An experimental method to measure the spin polarization of a ferromagnetic metal utilizing the Anomalous Hall effect was proposed and demonstrated. The merits of the method are its high precision (~0.1%) and its ability for measurement of a nano-sized object like a nanomagnet. The spin polarization of 81.7% and 89.4% was measured for the FeB and FeCoB nanomagnets, correspondingly. A few percent change of the spin polarization was observed under a gate voltage. The spin polarization linearly increases under a negative gate voltage and linearly decreases under a positive gate voltage. The dependence of the spin polarization on the polarity of the electrical current is observed. The polarity of the change depends on the asymmetry of the nanomagnet.